\documentclass[utf8]{FrontiersinHarvard} 

\usepackage{url,hyperref,lineno,microtype,subcaption}
\usepackage[onehalfspacing]{setspace}
\usepackage{siunitx}

\def\keyFont{\fontsize{8}{11}\helveticabold }
\def\firstAuthorLast{Lijuan Liu}
\def\Authors{Lijuan Liu\,$^{1,2,3,*}$}
\def\Address{$^{1}$Planetary Environmental and Astrobiological Research Laboratory (PEARL), School of Atmospheric Sciences, Sun Yat-sen University, Zhuhai, China \\
$^{2}$CAS center for Excellence in Comparative Planetology, China \\
$^{3}$Key Laboratory of Tropical Atmosphere-Ocean System, Sun Yat-sen University, Ministry of Education, Zhuhai, China}
\def\corrAuthor{Lijuan Liu}
\def\corrEmail{liulj8@mail.sysu.edu.cn}

\begin{document}
\onecolumn
\firstpage{1}

\title {The evolution of a spot-spot type solar active region which produced a major solar eruption} 

\author[\firstAuthorLast ]{\Authors} 
\address{} 
\correspondance{} 

\extraAuth{}

\maketitle

\begin{abstract}

\section{}

Solar active regions (ARs) are the main sources of large solar flares and coronal mass ejections. 
It is found that the ARs producing large eruptions usually show compact, highly-sheared polarity inversion lines (PILs). A scenario named as ``collisional-shearing'' is proposed to explain the formation of this type of PILs and the subsequent eruptions, which stresses the role of collision and shearing induced by relative motions of different bipoles in their emergence. However, in observations, if not considering the evolution stage of the ARs, 
about one third of the ARs that produce large solar eruptions govern a spot-spot type configuration. 
In this work, we studied the full evolution of an emerging AR, which owned a spot-spot type configuration when producing a major eruption,  
to explore the possible evolution gap between ``collisional shearing'' process in flux emergence and the formation of the spot-spot type, eruption-producing AR. We tracked the AR from the very beginning of its emergence until it produced the first major eruption. 
It was found that the AR was formed through three bipoles emerged sequentially. The bipoles were arranged in parallel on the photosphere, 
shown as two clusters of sunspots with opposite-sign polarities, 
so that the AR exhibited an overall large bipole configuration. 
In the fast emergence phase of the AR, the shearing gradually occurred due to the proper motions of the polarities, 
but no significant collision occurred due to the parallel arrangement of the bipoles. 
Nor did the large eruption occur. After the fast emergence phase, 
one large positive polarity started to show signs of decay. Its dispersion led to the collision
to a negative polarity which belonged to another bipole. A huge hot channel spanning the entire AR was formed through precursor flarings around the collision region. 
The hot channel erupted later, accompanied by an M7.3-class flare. 
The results suggest that in the spot-spot type AR, 
along with the shearing induced by the proper motions of the polarities, a decay process may lead to the collision of the polarities, driving the subsequent eruptions.  

\tiny
\keyFont{ \section{Keywords:} Solar active regions, Solar magnetic fields, Solar active region magnetic fields, Solar activity, Solar flares} 
\end{abstract}

\section{Introduction}


Solar flares and coronal mass ejections (CMEs) are the most violent phenomenon in the solar atmosphere,  
They both have the capability to cause hazardous space weather in the near-earth environment. 
Therefore, 
studying their origin are of fundamental importance in space weather forecasting. 
Except the ones associated with quiescent filament eruptions, 
most of the flares and CMEs originate from solar active regions (ARs). 
ARs appear as strong magnetic flux concentrations on the photosphere, which are believed to be formed through the magnetic flux emergence from the solar interior. The emergence of a flux tube may form the simplest bipolar region~\citep{Schmieder_2014}. 
The strong magnetic flux concentrations are manifested as regions containing dark sunspots in white light emission. 
The ability of the ARs to produce large eruptions varies greatly as suggested by the observations~\citep{Wang_2011c, Chen_Wang_2011}. 
To distinguish the productive and inert ARs, 
great efforts have been done in characterizing 
the magnetic properties of the ARs 
from the routine observations of the photospheric magnetic field, since the direct measurement of coronal magnetic field is not available up to now~\citep[e.g.,][]{Falconer_2002, Leka_Barnes_2003a, Falconer_etal_2006, Leka2007, Georgoulis_Rust_2007, Falconer_2008, Bobra_Couvidat_2015}. 
It is found that the ARs capable of producing large eruptions usually hold a more complex configuration~\citep[and references therein]{Toriumi_2016}, such as the $\delta$ sunspot where two umbrae of opposite-sign polarities sharing a common penumbra~\citep{Kunzel_1960}. 
This kind of complex configuration indicates a higher degree of non-potentiality, 
therefore the parameters characterizing the non-potentiality of the regions, such as magnetic shear, electric current density, current helicity, etc., tend to be higher in the region~\citep[e.g.,][]{Sunxd_2015, Lliu_2016}. 
It should be noted that the higher degree of non-potentiality, i.e., enough magnetic free energy is only a necessary condition to power large eruptions. The exact triggering of the eruptions might involve more specific mechanism such as magnetohydrodynamics (MHD) instabilities~\citep[e.g.,][]{Torok_2004, Kliem_2006} or magnetic reconnection~\citep[e.g.,][]{Antiochos_1999, Moore2001a}. 

When looking into the eruption-producing ARs, the strong-field, high-gradient sheared polarity inversion line (PIL), which is formed between opposite-sign polarities that are located in close proximity, is found to be a common source of large eruptions~\citep{schrijver2007characteristic, Schrijver_2009}. 
Shearing motion, sunspot rotation and converging motions are frequently observed near the sheared PIL~\citep[e.g.,][]{Zhang_2007, Li_2009, Green_2011, Yanxl_2015}. The former two are thought to be able to shear and twist the field lines, injecting magnetic free energy and magnetic helicity to the coronal field. 
The latter is believed to be able to bring opposite-sign magnetic field together, leading to flux cancellation (mild magnetic reconnection near the photosphere) which may form the magnetic flux rope, driving the subsequent eruptions as suggested by the flux cancellation model~\citep{VanBallegooijen_1989}. 

If considering the entire evolution of an AR from its emergence to decay, the eruptions may occur at any stage~\citep[and references therein]{Van_2015}. 
Nevertheless, observations suggest that compared to the ARs entering decaying phase, the ARs still under emerging and evolving tend to produce more violent eruptions~\citep{Schrijver_2009}. 

In general, it is suggested by the observations that the sheared PIL 
accompanied by the shearing motion and flux cancellation in emerging ARs is a place prone to major solar eruptions. 
The physics behind the phenomena is not entirely clear. Recently, a scenario name as ``collisional shearing'', 
is proposed to explain the origin of  large solar eruptions from emerging ARs~\citep{Chintzoglou_2019}. 
Through studying the evolution of two well-observed highly-productive ARs, the authors suggest that it is 
the relative motions between non-conjugated opposite-sign polarities leading to the formation of the sheared PIL and continuous shearing and flux cancellation around the PIL which are 
responsible to the series of eruptions. The relative motion between non-conjugated polarities is 
resulted from the proper separation between conjugated polarities when a flux tube emerges 
from the dense interior of the sun into the tenuous corona. 
Moreover, a statistical research on 19 ARs which emerged and produced at least one major eruption on the visible solar disk also suggests that the collisional shearing between non-conjugated polarities may be a common process at the source locations of large eruptions in emerging ARs~\citep{Liu_2021}. Those work stress the role of interaction between non-conjugated polarities of different bipoles in driving eruptions in emerging ARs. 

In observations, the emerging ARs evolve rapidly, 
so that the original conjugation of polarities may be blurred by multiple episodes of flux cancellation and magnetic reconnection later. Therefore, 
the most reliable way to determine the conjugation of polarities is to track the AR from the very beginning of its emergence to inspect the typical features between emerging conjugated polarities such as small moving dipoles on the photosphere~\citep{Strous_1999}, arch filaments in chromospheric observation, etc. However, the limited observations of photospheric magnetic field, which are mainly for the visible solar disk at present, restrict 
larger statistical research on the ``collisional shearing'' process. 
On the other hand,~\citet{Toriumi_2016} performed a statistical research on 29 ARs that produced 51 flares lager than GOES M5.0-class in solar cycle 24, 
mainly focusing on the pre-flare magnetic properties of the ARs rather than their evolution. 
In addition to the classification of magnetic patterns of $\delta$ sunspots proposed by ~\citet{Zirin_1987}, the authors classified the productive ARs into four groups, including the ``spot-spot'' group, ``spot-satellite'' group, ``quadrupole'' group,  and the ``inter-AR'' group. For the latter three groups, different bipoles are involved in forming the sheared PIL where the eruptions occur, so that the bipole-bipole interaction plays a role in driving the eruption.  
For the former group which accounts for about one third of the sample, the sheared PIL is formed between two major polarities or two groups of opposite-sign sunspots, it is not clear if the bipole-bipole interaction and thus ``collisional shearing'' exists and plays a role in driving eruptions. 
Further study is needed. 

In this work, we searched the sample in~\citet{Toriumi_2016}, and found an AR (NOAA AR 12036) 
of spot-spot configuration which emerged and produced 
an M7.3-class eruptive flare on the visible solar disk. We tracked the evolution of the AR to see what drove it to produce the large eruption. 

\section{Data analysis} 

We used the photospheric vector magnetograms 
provided by the Helioseismic and Magnetic Imager~\citep[HMI,][]{Scherrer_2012} on-board the Solar Dynamics Observatory~\citep[SDO,][]{Pesnell_2012} to investigate the evolution of the magnetic properties of the AR. The data has a plate scale of 0.$''$5 and a temporal cadence of 720~s. Here a data segment which is de-projected from the native helioprojective Cartesian coordinate to a cylindrical equal area (CEA) coordinate is used~\citep{Bobra_2014}. HMI also provides the photospheric continuum intensity maps, which is used here to check the evolution of the sunspots in the AR. 

We used the ultraviolet (UV) and extreme-ultraviolet (EUV) images provided by the Atmospheric Imaging Assembly~\citep[AIA,][]{Lemen2012} on-board the SDO to check the coronal evolution of the AR and the eruption details of the M7.3-class flare. The data has a plate-scale of 0.$''$6 and a temporal cadence up to 12~s. 

To determine the connectivity between conjugated polarities in the early emergence phase more accurately, we performed a potential field extrapolation~\citep{Alissandrakis_1981} to the normal component of the photospheric magnetic field ($B_z$) and analyzed the extrapolated field lines. 


We analyzed the motions and flows in the ARs in two ways. The first way is to track the macroscopic motion of the large polarities which formed the main PIL where the eruption occurred. 
The second is to analyze the microscopic flow fields of the region. For the first way, we firstly detected the local peaks in each polarities on each magnetogram of $B_z$, and then for each polarity, we calculated a flux-weighted centroids in a circle which took the detected peak as the center and 
a radius of 5 Mm~\citep{Chintzoglou_2019, Liu_2021}. For the second way, we used a Fourier-based local correlation tracking (FLCT) method~\citep{Welsch_2004} to calculate the horizontal velocity field of the intensity features on the $B_z$ magnetograms. 
The velocity field is suggested to be able to affect the photospheric footpoints of the coronal field.  
 
We extracted the source PIL of the eruption through drawing contour line of $B_z=0$ on the smoothed $B_z$ magnetograms. 
To further assess the degree of collision along the PIL, we calculated the gradient of $B_z$ across the PIL through $\triangledown B_z=\sqrt{{(\frac{\partial B_z}{\partial x})}^2+{(\frac{\partial B_z}{\partial y})}^2}$.

\section{Results}

\subsection{Evolution of the AR}

\begin{figure}[h!]
\begin{center}
\includegraphics[width=16cm]{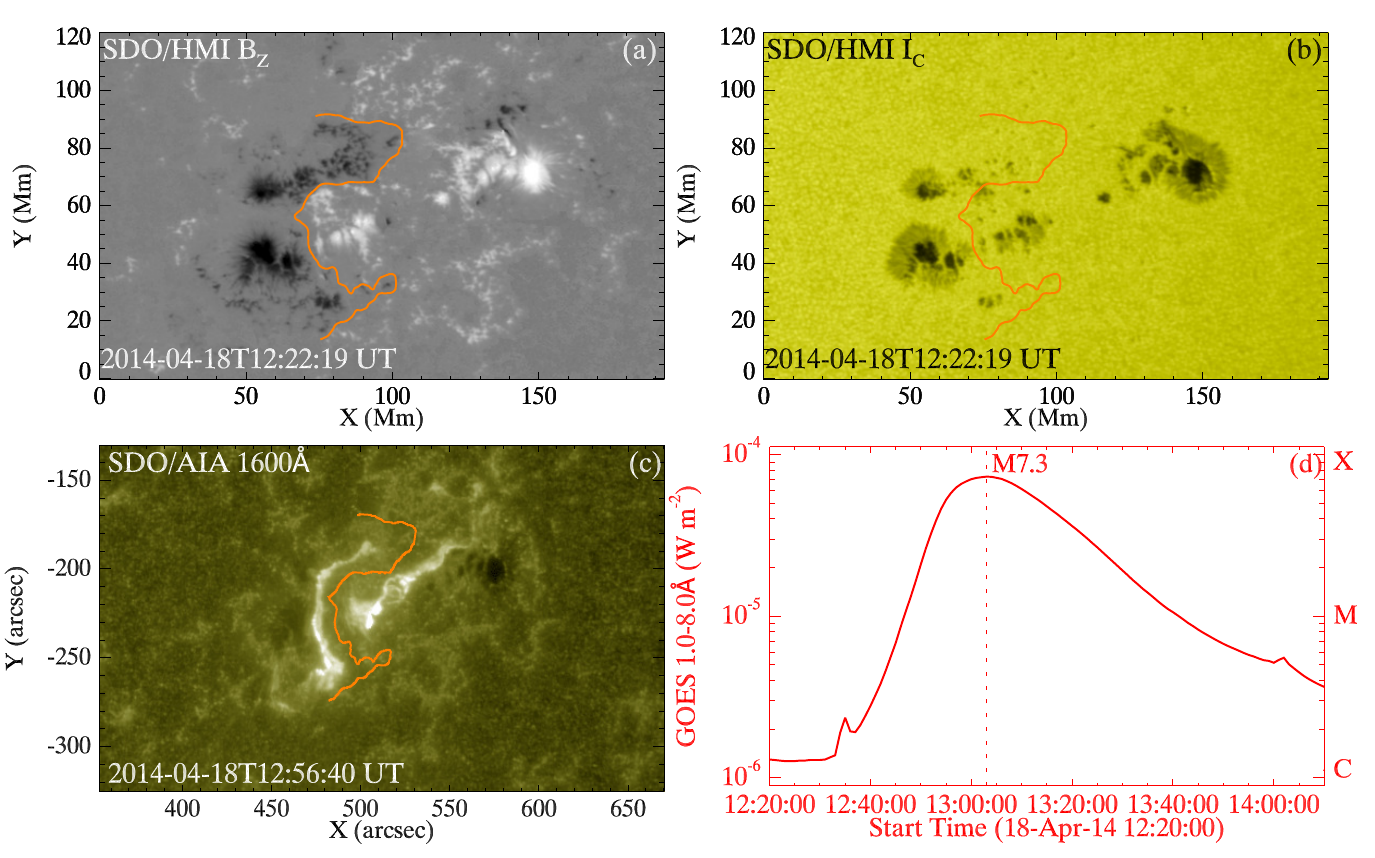}
\end{center}
\caption{Overview of NOAA AR 12036 and its first major eruption, an M7.3-Class flare (SOL2014-04-18T12:31). 
(a) The photospheric $B_z$ map prior to the flare, in which $B_z$ saturates at $\pm$ 2000 Gauss. (b) The photospheric continuum intensity at the same time as panel (a). (c) The SDO/AIA 1600~\AA~images taken during the flare. The orange curves in the three panels are the PILs where the eruption occur. 
(d) The light curve of the GOES 1-8~\AA~flux during the M7.3-class flare. The vertical line indicates the peak time of the flare. }\label{fig:1}
\end{figure}

The AR (NOAA AR 12036) started to emerge from around 2014-04-12 03:22 UT in the southern hemisphere.  
Except a series of small flarings, it produced no major eruption (flares larger than M-class or CMEs) until 18 April. The first major eruption from the AR is an M7.3 class flare (SOL2014-04-18T12:31). 
The AR held an overall bipolar configuration, i.e., a spot-spot configuration prior to the flare, consisting of two groups of sunspots of opposite-signed flux (Figures~\ref{fig:1}(a)-(b)). 
The flare originated from the main PIL formed between the two groups of polarities 
as suggested by the flare ribbons 
along the PIL (Figure~\ref{fig:1}(c)). 

\begin{figure}[h!]
\begin{center}
\includegraphics[width=16cm]{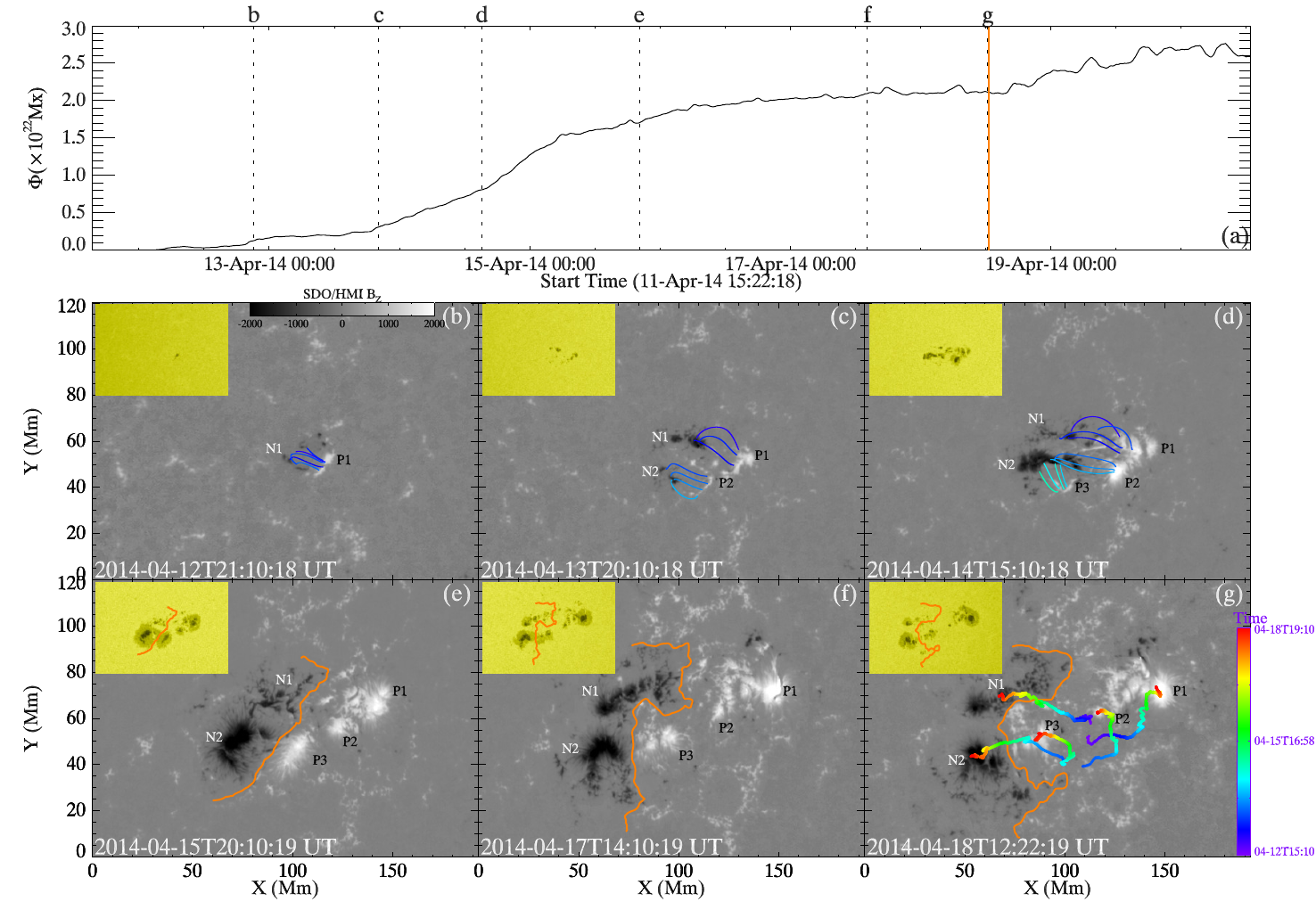}
\end{center}
\caption{Overall evolution of the AR. (a) The unsigned magnetic flux $\Phi$ of the AR. The vertical-dashed lines indicate the times of panels (b)-(g). The orange vertical line marks the start time of the M7.3-class flare. (b)-(g) Snapshots of the photospheric $B_z$ showing the evolution of the AR. The insets in each panel display the photospheric continuum intensity at the same time as $B_z$ maps. The labels ``P1'', ``N1'', ``P2'', ``N2'', and ``P3'' mark different polarities. The colored lines in panels (b)-(d) are the coronal field lines extracted from the potential field extrapolation, showing the connectivity of different polarities in their early emergence phase. The colored curves in panel (g) display the trajectories of the centroids of each polarity, of which the color is coded by the elapsed time. The orange curves in (e)-(g) are the source PIL 
extracted on magnetograms at different times.}\label{fig:2}
\end{figure}

We firstly checked the unsigned magnetic flux ($\Phi$) of the entire AR. In general, the $\Phi$ increased fast before around 2014-04-15 09:10 UT, reaching $1.55\times 10^{22} $ Mx with a mean emergence rate of $1.98\times 10^{20} $ Mx per hour. 
The increase 
then slowed down, as the $\Phi$ reached $2.11\times 10^{22} $ Mx before the flare 
with a mean emergence rate of $7.50\times 10^{19} $ Mx per hour. 
Through checking the photospheric magnetograms, we found that the AR was roughly formed by three bipoles which emerged sequentially.  
The first bipole (polarities P1 and N1 in Figure~\ref{fig:2}), the appearance of which made the AR appear, 
started to emerge from around 2014-04-12 03:22 UT. 
It emerged in the northern region of the AR if we took the pre-flare configuration of the AR (Figure~\ref{fig:1}(a)) as the reference.   
The second bipole (polarities P2 and N2 in Figure~\ref{fig:2}) started to emerge from around 2014-04-13 18:10 UT in the southern region of the AR. 
The two bipoles were located parallel, i.e., the axis connecting the two main conjugated polarities of each bipole was parallel to each other. 
From around 2014-04-14 09:10 UT, a positive polarity (marked as P3 in Figure~\ref{fig:2}) started to emerge 
in between the two polarities of the second bipole.   
The negative flux connected to P3 was part of the polarity N2 and therefore can not be separated from N2. 
The connectivity 
between the conjugated polarities of each bipole in their early emergence was confirmed by the potential field extrapolation (colored lines in Figures~\ref{fig:2}(b)-(d)). 
Although the three bipoles all started to emerge 
before 2014-04-15 09:10 UT, i.e., 
in the overall 
fast emergence phase we described above, 
the appearance of each bipole 
affected the emergence rate differently. 
Before the second bipole appeared, 
the magnetic flux increased slowly  
with a mean emergence rate of around $6.18\times 10^{19} $ Mx per hour.  
After the appearance of the second bipole, the $\Phi$ increased faster with a mean emergence rate of around $2.60\times 10^{20} $ Mx per hour. 
After the third bipole appeared, 
the $\Phi$ increased even more faster with a mean emergence rate of around $3.82\times 10^{20} $ Mx per hour until 2014-04-15 09:10 UT, after which the flux emergence slowed down. 

The trajectories of the polarities (colored lines in Figure~\ref{fig:2}(g)) 
showed the separation between conjugated polarities, and an overall northward motion of all polarities. 
As emergence went on, 
all of the positive polarities were located on one side, and all of the negative polarities 
were located on the other side, so that a single 
PIL can be drawn between them (orange curves in Figures~\ref{fig:2}(e)-(g)). 
The PIL obtained before the flare (Figures~\ref{fig:2}(g)) was more sheared than that obtained 
earlier (Figures~\ref{fig:2}(e)). 
This may be because that the proper motions of different polarities 
were different in magnitudes and directions, so that the PIL may be sheared gradually. 
For example, for the polarities P3 and N1, although both polarities showed northeastward motion, 
N1 passed through a longer trajectory than P3 in the same period (Figure~\ref{fig:2}(g)), 
therefore resulting in a shear motion of N1 relative to P3. 
The M7.3-class flare occurred above 
the sheared PIL eventually. 

\subsection{Evolution of the polarity P3}

\begin{figure}[h!]
\begin{center}
\includegraphics[width=16cm]{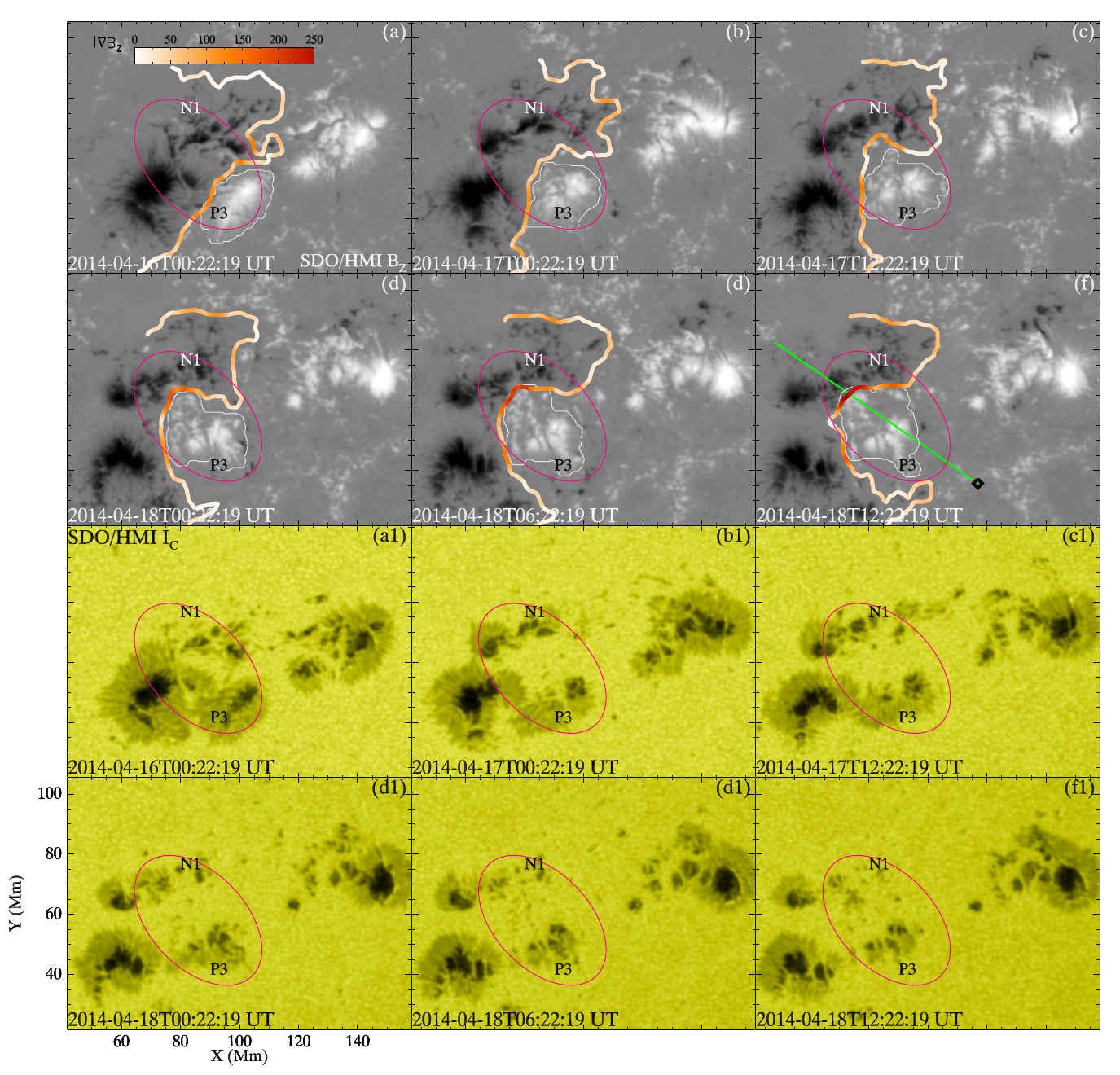}
\end{center}
\caption{The evolution of the positive polarity P3. (a)-(f) Snapshots of the $B_z$ magnetograms zoomed on the polarities P3 and N1. The white contours outline the boundary of P3, within which the magnetic flux of P3 is calculated. The green line in panel (f) indicates the slice (taking the black diamond symbol as the start point) used to generate the time-distance plot of $B_z$ in Figure~\ref{fig:add}(a). (a1)-(f1) The continuum intensity images corresponding to the $B_z$ maps. The magenta ellipse in each panel marks the interface region of the polarities P3 and N1. The colored curves  
are the PILs at each moment, the color of which is coded by the gradient of $B_z$ ($\triangledown {B_z}$, in unit of G Mm$^{-1}$) across the PIL.}\label{fig:3}
\end{figure}

\begin{figure}[h!]
\begin{center}
\includegraphics[width=16cm]{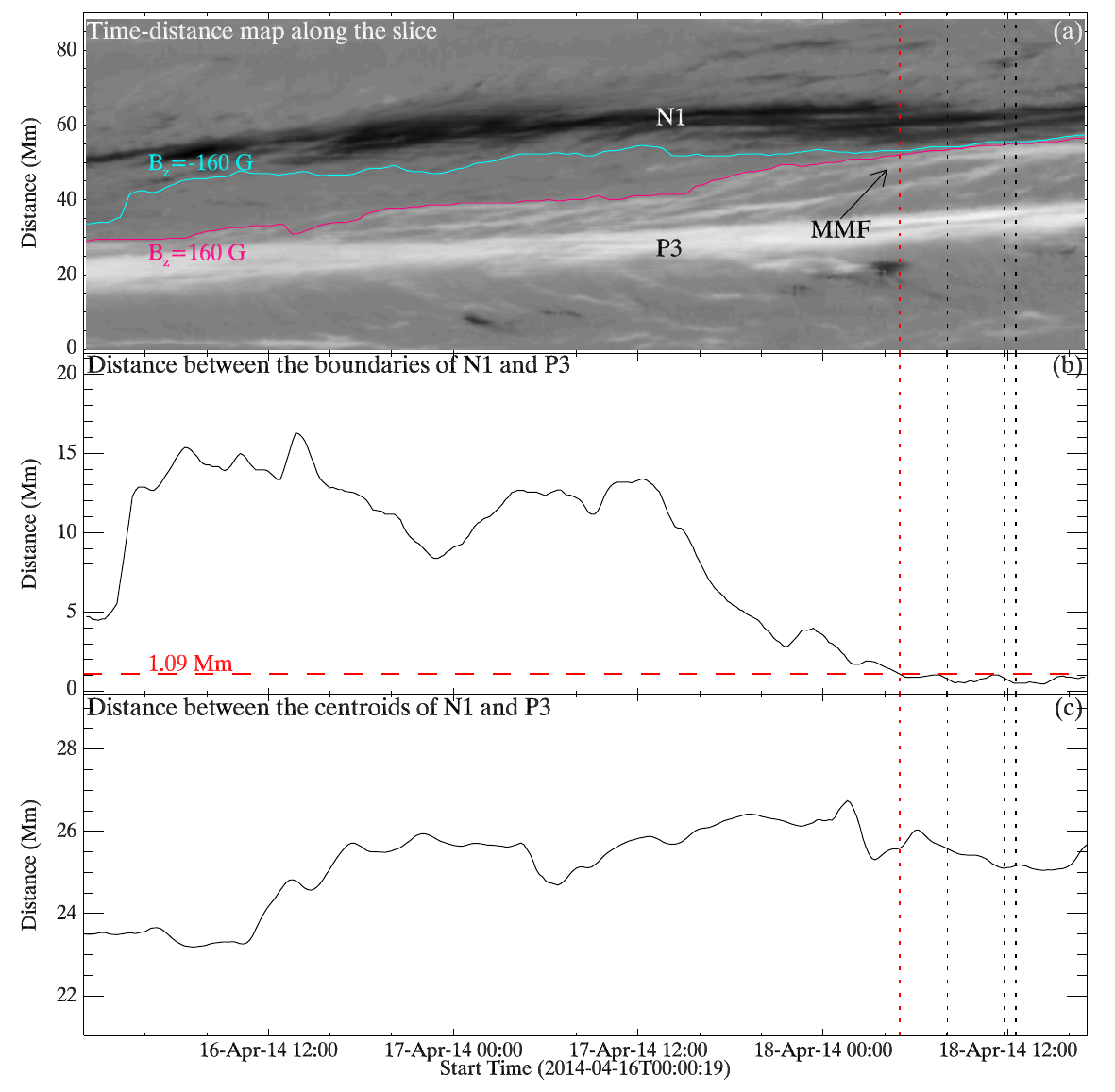}
\end{center}
\caption{(a) The time-distance map of $B_z$ along the slice (shown as a green line Figure~\ref{fig:3}(f)). The cyan and magenta curves are the contour lines of -160 Gauss and 160 Gauss draw from the map, respectively, outlining the inner boundaries of the polarities N1 and P3. The black arrow indicates the moving magnetic features streaming 
away from P3. The red vertical line marks the time when the distance between the inner boundaries of polarities P3 and N1 drops below the width of 3 pixels (1.09 Mm). The three black vertical lines indicate the start timings of the first precursor flaring, the second precursor flaring and the M7.3-class flare, respectively. (b) The distance between the inner boundaries of the polarities P3 and N1. The red horizontal line indicates the distance value of 1.09 Mm. (c) The distance between the flux-weighted centroids of the polarities P3 and N1. The vertical lines in panels (b) and (c) have the same meanings as the ones in panel (a). }\label{fig:add}
\end{figure}

\begin{figure}[h!]
\begin{center}
\includegraphics[width=16cm]{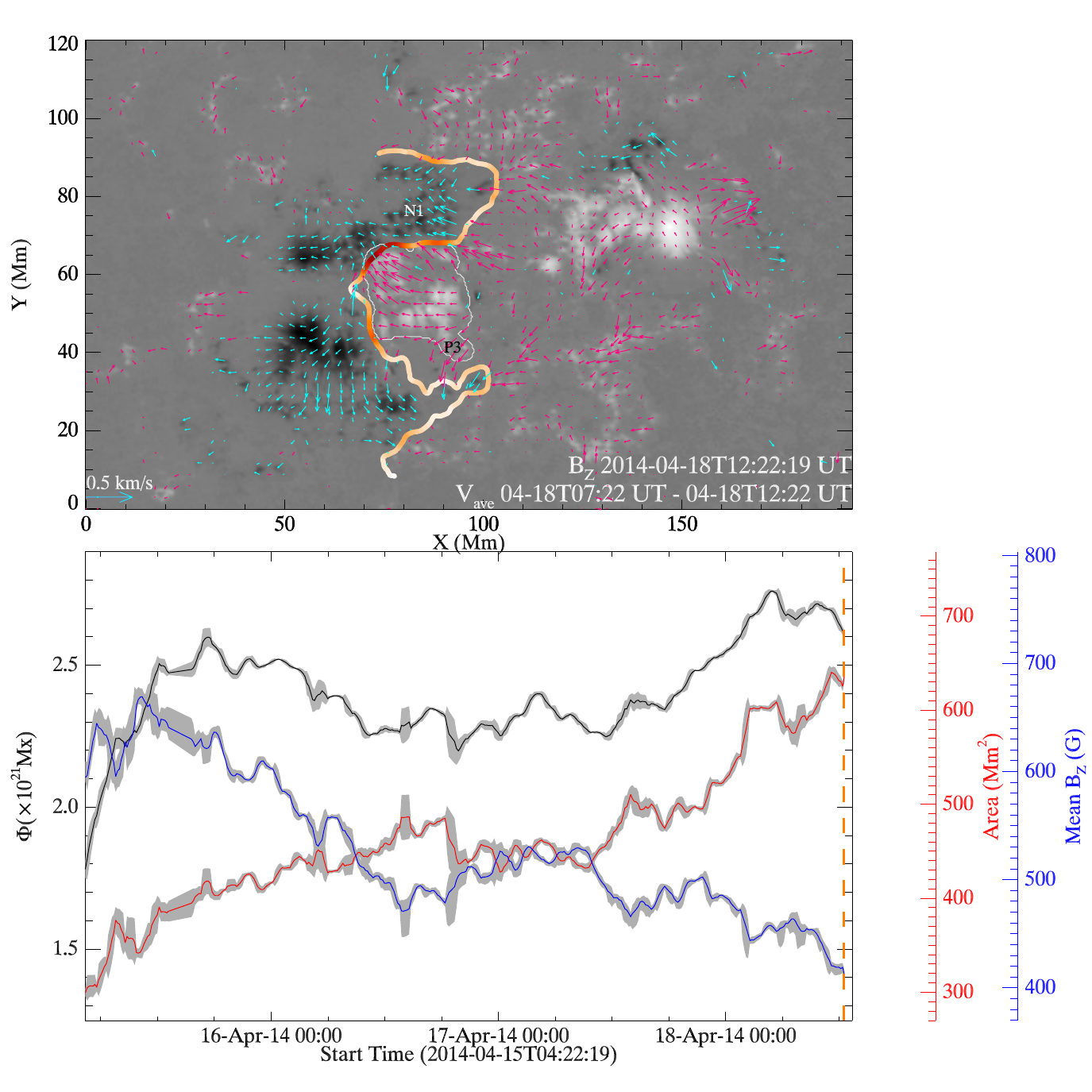}
\end{center}
\caption{(a) The horizontal velocity field ($V_{ave}$) averaged in five hours prior to the M7.3 class flare. The magenta arrows are for the positive flux, while the cyan arrows are for the negative flux. The PIL and boundary of polarity P3 have the same meaning as the ones in Figure~\ref{fig:3}. (b) Unsigned magnetic flux, area and mean $B_z$ of polarity P3. 
The grey regions mark the uncertainties of the parameters. The orange vertical line marks the start time of the M7.3-class flare.}\label{fig:4}
\end{figure}

Before the flare, 
the middle part of the PIL was 
formed between the non-conjugated P3 and N1 (Figure~\ref{fig:2}(g)), 
showing a sign of collision, 
i.e., the two polarities were located very close to each other. Through tracking the two polarities, 
we found that the collision was mainly resulted from the evolution of the polarity P3. The results are shown in  
Figures~\ref{fig:3}-\ref{fig:4}. 

In the early phase of the emergence, 
the polarities P3 and N1 were not that close to each other, 
as indicated by the relatively low gradient of $B_z$ ($\triangledown B_z$) 
across the PIL (colorded curves in Figures~\ref{fig:3} (a)-(b)). 
Later, although the flux emergence of the entire AR continued,  
the polarity P3 showed a sign of decay: 
its area became larger; the sunspot in it lost the coherence gradually, and became a set of pores; 
small magnetic features were found streaming away 
from the polarity P3, moving to N1 (Figures~\ref{fig:3}(c)-(f) and Figures~\ref{fig:3}(c1)-(f1)). 
They are usually called as moving magnetic features ~\citep[MMFs,][]{Van_2015}. 
As the MMFs approached the boundary of N1, the collision gradually occurred at the interface between P3 and N1.  
This was supported by the growing  
$\triangledown B_z$ in the PIL part between the polarities P3 and N1 (darker color of PILs in Figures~\ref{fig:3} (d)-(f)). 

The above process was further checked in detail through analyzing the time-distance map of $B_z$ in a slice (green line in Figure~\ref{fig:3} (f)) across the interface between the polarities P3 and N1 (Figure~\ref{fig:add}). The inner boundaries of P3 and N1, i.e., the boundary parts next to the PIL, were extracted from the time-distance map with the thresholds of -160 Gauss and 160 Gauss, respectively (cyan and magenta lines in Figure~\ref{fig:add}(a)). Through measuring the distance between the boundaries of the two polarities, 
we found that the distance firstly increased from around 4.70 Mm to 16.28 Mm in the early emergence phase, and then varied between 13.36 Mm and 16.28 Mm until 2012-04-17 12:10 UT (Figure~\ref{fig:add}(b)). It then decreased fast, dropping below 1.09 Mm from around 2014-04-18 04:58 UT, and varied between 0.48 to 1.09 Mm until the M7.3-class flare. The value of 1.09 Mm is the width of three pixels on the magnetograms. If taking this value as the threshold of collision, one can conclude that the collision occurred from around 2014-04-18 04:58 UT. The fast approaching between P3 and N1 was mainly resulted from the MMFs (indicated by the black arrow in Figure~\ref{fig:add}(a)). 
We also checked the distance between the flux-weighted centroids of the two polarities (Figure~\ref{fig:add}(c)). It is seen that this distance did not change much. It firstly increased from 23.50 Mm to 26.74 Mm, and then slightly decreased to 25.15 Mm prior to the M7.3-class flare. The results suggest that the collision was mainly resulted from the MMFs 
streaming away from P3 rather than the overall large-scale motions of the polarities.  
The flows described above 
were also checked 
by the flow field in the polarity P3: 
the velocities averaged in five hours before the flares confirmed 
a northeastward flow heading to the polarity N1 (Figure~\ref{fig:4}(a)). 

We further analyzed the evolution of the polarity P3 quantitatively. The boundary of the polarity P3 on each magnetogram was determined through a series of morphological operation, 
including the morphological open, morphological close, and region growing, which were performed on the strong field kernels selected by a threshold of 160~G. 
The slight discrepancy between the boundary of P3 and the PIL was because that the former was extracted from the original magnetogram while the later was extracted from the smoothed magnetogram.  
The unsigned magnetic flux $\Phi$, area, and mean $B_z$ of the polarity P3 were then calculated (Figure~\ref{fig:4}(b)). It is found that the unsigned magnetic flux of the polarity (black curve in Figure~\ref{fig:4}(b)) increased fast before around 
2013-04-15 17:22 UT to around $2.60\times 10^{21}$ Mx. 
After the fast emergence phase, the $\Phi$ varied slowly between $2.25\times 10^{21}$ to $2.60\times 10^{21}$ Mx 
until 2013-04-17 18:10 UT. 
A slow emergence phase occurred later, 
within which the $\Phi$ increased to around $2.70\times 10^{21}$ Mx before the flare. 
The area of the polarity (red curve in Figure~\ref{fig:4}(b)) kept increasing, 
reaching around 638~Mm$^2$ before the flare. 
The mean $B_z$ of the region (blue curve in Figure~\ref{fig:4}(b)) increased to 670 Gauss in the fast emergence phase before around 2013-04-15 12:10 UT, 
and then kept decreasing to 420 Gauss 
before the flare. 
The evolution of the three parameters suggested that after the fast emergence phase, although the slow flux emergence still continued,  
the polarity P3 underwent a significant dispersion as suggested by the increasing area and decreasing mean $B_z$ in it. Combined with the MMFs and decaying sunspot in P3 observed in the magnetograms and continuum intensity images (Figure~\ref{fig:3}), we concluded that the polarity P3 underwent a decaying process in general.

\subsection{Precursor flarings and the M7.3-class flare}

\begin{figure}[h!]
\begin{center}
\includegraphics[width=16cm]{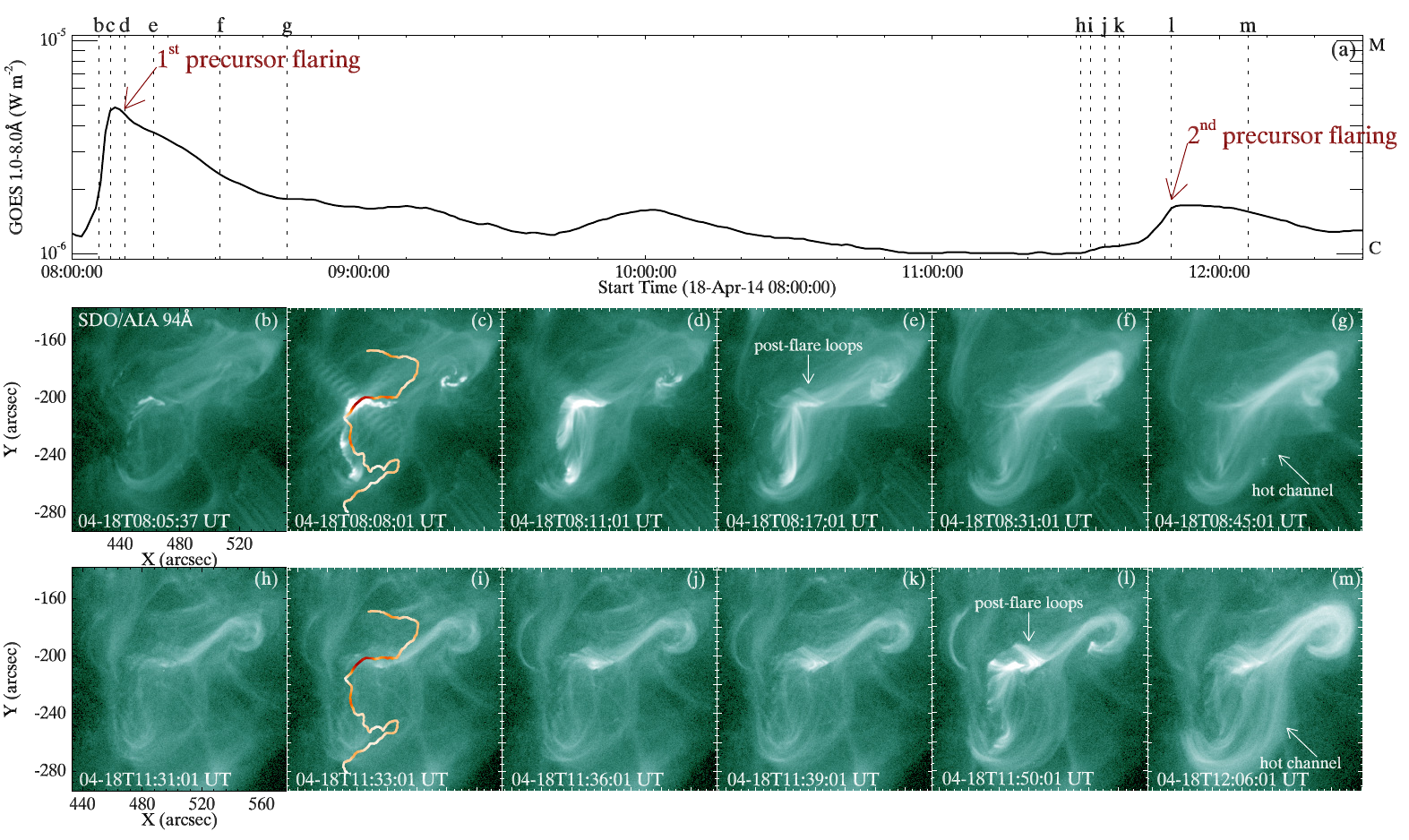}
\end{center}
\caption{Two precursor small flarings prior to the M7.3-class flare. (a) The light curve of the GOES 1-8~\AA~flux during the two precursor flarings. The vertical-dashed lines mark the time of the EUV images shown below. (b)-(g) SDO/AIA 94~\AA~images capturing the first precursor flaring. (h)-(m) SDO/AIA 94~\AA~images for the second precursor flaring. The colored curves 
in (c) and (i) are PILs, the color of which has the same meaning as the ones in Figure~\ref{fig:3}.}\label{fig:5}
\end{figure}

\begin{figure}[h!]
\begin{center}
\includegraphics[width=16cm]{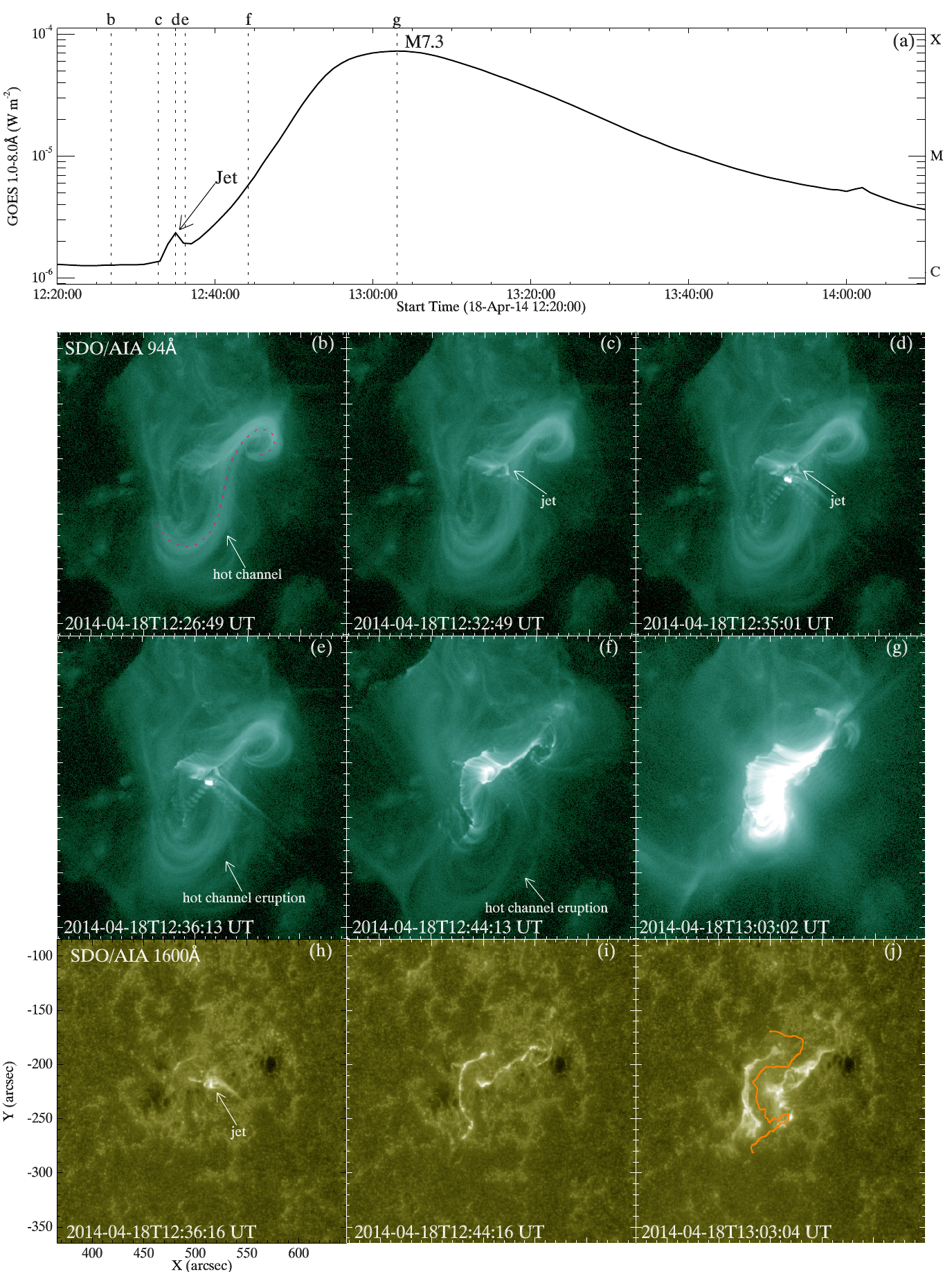}
\end{center}
\caption{The eruption details of the M7.3-class flare. (a) The light curve of the GOES 1-8~\AA~flux during the flare. (b)-(g) Snapshots of SDO/AIA 94~\AA~images during the flare. (h)-(j) Snapshots of SDO/AIA 1600~\AA~images during the flare. The orange curve outlines the source PIL of the eruption. 
}\label{fig:6}
\end{figure}

The above results suggested that the decay of polarity P3 after the fast emergence phase 
resulted in 
the collision to polarity N1, creating a location favourable of flux cancellation and reconnection. 
We further checked the evolution of the AR in the EUV observations, and found that before the flare, a hot channel which is deemed 
as the proxy of a flux rope was formed through small-scale flarings ignited near the interface between P3 and N1. The eruption of the hot channel was responsible to the M7.3-class flare and an accompanied CME (not shown here). 
The results are shown in Figure~\ref{fig:5} and Figure~\ref{fig:6}. 

In 5 hours before the major flare, 
two main precursor flarings occurred. The first one is recorded as a C4.8-class flare (SOL2014-04-18T08:03). 
In this precursor, the flaring ignited 
from the high gradient part of the PIL, i.e., the collisional interface between P3 and N1 (Figures~\ref{fig:5}(b)-(g)). As the flaring went on, small sets of post-flare loops 
appeared above the high gradient part of the PIL, accompanied by the appearance of a hot channel in the SDO/AIA 94~\AA~images. These suggested that there were tether-cutting type reconnection occurred around the interface region between the polarities P3 and N1, 
i.e., in between the sheared loops connecting P3 and N2 and those connecting P1 and N1, forming the nearly-potential 
post-flare loops below and the continuous 
hot channel above. The second precursor flaring (started from around 2014-04-18 11:44 UT) was similar to the first one. It was manifested as a mild bump in the GOES soft X-ray flux and therefore 
was not being classified as a GOES flare. However, the SDO/AIA 94~\AA~images suggested that the flaring also ignited from the interface between polarities P3 and N1, after which the post-flare loops 
appeared above the collisional PIL part, accompanied by the appearance of a continuous S-shaped hot channel spanning the entire PIL (Figures~\ref{fig:5}(h)-(m)). 

After the second precursor, the hot channel was visible in the 94~\AA~passbands (Figure~\ref{fig:6}(b)) until the M7.3-class flare started. 
In the very early phase of the flare, 
another small-scale flaring occurred, manifested as a small peak in the GOES 1-8~\AA~flux (indicated by an arrow in Figure~\ref{fig:6}(a)). 
The small flaring was a jet process occurring at a possible null point as suggested by the cusp-shaped structure and mass flow in the 94~\AA~passbands images (Figures~\ref{fig:6}(b)-(d)). 
The jet mass flow was even visible in the AIA 1600~\AA~passband. 
After the small flaring, 
the hot channel started to rise and finally erupted out (Figures~\ref{fig:6}(e)-(g)). 
From the EUV image (Figure~\ref{fig:6}(c)) we can estimate that 
when the jet occurred, 
the distance between the two footpoints of the cusp-shaped structure was around 7 Mm, while the distance between the two sets of footpoints of the post-flare loops 
formed in precursor flarings was around 18 Mm. If taking the distances as rough indicators to the heights of the coronal structures, the results suggested that the jet was more likely to occur lower than the loops, i.e., lower than the hot channel formed above the loops during the precursor flarings.  
In general, the jet process seemed to serve as 
a trigger of the hot channel eruption rather than a process forming or enhancing the channel~\citep{Liu_2018d}. 

\section{Summary and conclusion}

In this work, we tracked the evolution of the NOAA AR 12036 which emerged and produced an M7.3-class flare on the visible solar disk. It held a spot-spot type configuration before the flare, i.e., a monolithic PIL was formed between the group of positive polarities and the group of negative polarities, from which the M7.3-class flare occurred. 
The observations suggested that the AR was formed by three bipoles that emerged sequentially. The two bipoles emerging earlier were located parallel. The third bipole emerged in between the two polarities of the second bipole. Its negative polarity was mixed with 
the negative polarity of the second bipole.  
In the early phase when the flux emergence was fast, the $B_z$ gradient across the PIL remained small, and no collision or large eruption occurred.

Later, although the slow flux emergence of the entire AR continued, the positive polarity P3 of the third bipole started to show signs 
of decay: 
the sunspot in P3 lost its coherence gradually;   
small magnetic features were found stripped 
away from the polarity, 
heading toward the negative polarity N1 of the bipole emerged firstly. 
The increasing area and decreasing mean $B_z$ of the polarity P3 further supported that P3 was undergoing a decaying process.  
The small moving magnetic features brought the polarities P3 and N1 into contact about 7.5 hours prior to the flare, i.e., the collision occurred between the polarities 
since then.  
The $B_z$ gradient at the interface between P3 and N1 increased with the converging and collision process. 

Accompanied by the collision between P3 and N1, two precursor flarings occurred in five hours before the flare, both of which ignited from around the interface region between P3 and N1. 
They all formed post-flare loops above the interface region and ignited a continuous S-shaped hot channel above the loops. 
In the very early phase of the M7.3-class flare, 
another small-scale flaring occurred. 
It was a jet process occurring at a possible null point located lower than the hot channel, and may disturb the hot channel to erupt later.  
The hot channel eruption induced the main phase flaring of the 
flare and a CME. 


Combining the above results, we found that it was the collision between the non-conjugated polarities in the decay phase of one polarity created a location favorable of cancellation and flaring. 
The collision brought the footpoints of different sets of sheared loops belonging to different bipoles into close proximity, 
enabling the tether-cutting type reconnection to occur between them. 
The resultant pre-flare flarings and possible cancellation helped to form and enhance the flux rope which was manifested as an S-shaped hot channel. 
The eruption of the hot-channel was directly triggered by a low-lying jet process, similar as the case reported in ~\citet{Liu_2015}, which showed a jet process triggering a CME. Note that accompanied by the collision, shearing process is also needed to create the helical field lines of the flux rope. Similar as in the rapidly emerging non-bipolar ARs~\citep{Chintzoglou_2019}, the shearing process in the spot-spot type AR here was also induced by the proper motions between non-conjugated polarities. The proper motions of the two non-conjugated polarities before the collision had similar direction but different magnitudes, so that the shearing gradually occurred between them.  

The ``collisional shearing'' reported in \citet{Chintzoglou_2019} and \citet{Liu_2021} mainly 
occurred between the non-conjugated polarities in emerging non-bipolar ARs, and 
was resulted from the relative motions of non-conjugated polarities. Those motions were indeed caused by the proper separation 
of conjugated polarities. In the NOAA AR 12036 studied here, 
the arrangement of the bipoles, which was an overall bipolar configuration, did not create a condition favourable of contact and collision between 
non-conjugated polarities 
in the fast emergence phase, 
therefore no 
large eruption occurred in this phase. 
Later on, 
although slow flux emergence continued, one of the main positive polarities started to decay,  
leading to collision between the non-conjugated polarities 
eventually. 
The two precursor flarings which helped to form the hot channel that erupted later 
were the result of collision. 


Although the photospheric collision plays an important role in most ARs which produce the large solar eruption, it is not an absolute necessary condition to eruption. In a few events like the ``inter-AR'' type cases in ~\citet{Toriumi_2016}, which accounts about $7\%$ of all their sample, 
the eruption occurred above a PIL without close contact or strong collision between opposite-sign polarities. They may be resulted from the process occurring in the corona, the details behind which need further study. 

The results in this work may explain why there is no single bipolar AR in the statistical research of ``collisional shearing''~\citep{Liu_2021}, though one should still be aware that their sample size is limited.  
For the AR holding a non-bipolar configuration, 
it contains more than one bipole when emerging. 
The arrangement of the bipoles 
may be prone to collisional shearing to occur between non-conjugated opposite-sign polarities. 
The flux emergence drives 
the collisional shearing to continue, 
eventually leading to eruptions. 
For the ``spot-spot'' type AR we studied here, 
the bipoles were roughly parallel to each other in the fast emergence phase, the arrangement of which did not create a condition favourable of collision. Although the shearing gradually occurred between the non-conjugated polarities due to their proper motions, 
no large eruption occurred until the collision occurred as a result of 
the polarity decay. 
Note that we have only one case here, it is possible that in some other ``spot-spot'' type ARs, the ``collisional shearing'' process may not even occur. The rotation and shearing of the sunspots driven by other process may play a major role in producing large eruptions in those ARs as suggested by the previous research \citep{Zhang_2007, Yan_2007, Yan_2008, Li_2009, Yan_2009}. 
Statistical research is needed to check if ``collisional shearing'' is common in the ``spot-spot'' type ARs.  
On the other hand, 
even in the ``spot-spot'' type AR here, the collision occurred between non-conjugated polarities at the source PIL of the eruption, not at the self PIL between conjugated polarities. Statistical research considering the different evolution stages of the ARs is also needed to check whether this is common.

\section*{Author Contributions}

Lijuan Liu conducted the analysis of the data, drafted and completed the manuscript. 

\section*{Funding}
Lijuan Liu acknowledges the support received from the National Natural Science Foundation of China
(NSFC grant no. 12273123), 
and from the Guangdong Basic and Applied Basic Research Foundation (2022A1515011548).

\section*{Acknowledgments}
We acknowledge the SDO and GOES missions for providing quality observations. 

\section*{Supplemental Data}
Video 1 | Online movie 1, generated from a sequence of images similar as in Figure~\ref{fig:2}. \\
Video 2 | Online movie 2, generated from a sequence of images similar as in Figure~\ref{fig:3}. \\
Video 3 | Online movie 3, generated from a sequence of images similar as in Figure~\ref{fig:5}. \\
Video 4 | Online movie 4, generated from a sequence of images similar as in Figure~\ref{fig:6}.


\bibliographystyle{Frontiers-Harvard} 
\bibliography{Spot-spot_AR}

\begin{thebibliography}{41}
\providecommand{\natexlab}[1]{#1}
\expandafter\ifx\csname urlstyle\endcsname\relax
  \providecommand{\doi}[1]{doi:\discretionary{}{}{}#1}\else
  \providecommand{\doi}{doi:\discretionary{}{}{}\begingroup
  \urlstyle{rm}\Url}\fi
\providecommand{\selectlanguage}[1]{\relax}
\providecommand{\bibAnnoteFile}[1]{%
  \IfFileExists{#1}{\begin{quotation}\noindent\textsc{Key:} #1\\
  \textsc{Annotation:}\ \input{#1}\end{quotation}}{}}
\providecommand{\bibAnnote}[2]{%
  \begin{quotation}\noindent\textsc{Key:} #1\\
  \textsc{Annotation:}\ #2\end{quotation}}

\bibitem[{Alissandrakis(1981)}]{Alissandrakis_1981}
Alissandrakis, C.~E. (1981).
\newblock {On the computation of constant $\alpha$ force-free magnetic field}.
\newblock \emph{A{\&}A} 100, 197--200
\bibAnnoteFile{Alissandrakis_1981}

\bibitem[{Antiochos et~al.(1999)Antiochos, DeVore, and
  Klimchuk}]{Antiochos_1999}
Antiochos, S.~K., DeVore, C.~R., and Klimchuk, J.~A. (1999).
\newblock {A Model for Solar Coronal Mass Ejections}.
\newblock \emph{ApJ} 510, 485--493.
\newblock \doi{10.1086/306563}
\bibAnnoteFile{Antiochos_1999}

\bibitem[{Bobra and Couvidat(2015)}]{Bobra_Couvidat_2015}
Bobra, M.~G. and Couvidat, S. (2015).
\newblock {Solar Flare Prediction Using SDO /HMI Vector Magnetic Field Data
  With a Machine-Learning Algorithm}.
\newblock \emph{ApJ} 798, 135.
\newblock \doi{10.1088/0004-637X/798/2/135}
\bibAnnoteFile{Bobra_Couvidat_2015}

\bibitem[{Bobra et~al.(2014)Bobra, Sun, Hoeksema, Turmon, Liu, Hayashi
  et~al.}]{Bobra_2014}
Bobra, M.~G., Sun, X., Hoeksema, J.~T., Turmon, M., Liu, Y., Hayashi, K.,
  et~al. (2014).
\newblock {The Helioseismic and Magnetic Imager (HMI) Vector Magnetic Field
  Pipeline: SHARPs – Space-Weather HMI Active Region Patches}.
\newblock \emph{SoPh} 289, 3549--3578.
\newblock \doi{10.1007/s11207-014-0529-3}
\bibAnnoteFile{Bobra_2014}

\bibitem[{Chen et~al.(2011)Chen, Wang, Shen, Ye, Zhang, and
  Wang}]{Chen_Wang_2011}
Chen, C., Wang, Y., Shen, C., Ye, P., Zhang, J., and Wang, S. (2011).
\newblock {Statistical study of coronal mass ejection source locations: 2. Role
  of active regions in CME production}.
\newblock \emph{JGRA} 116, A12108.
\newblock \doi{10.1029/2011JA016844}
\bibAnnoteFile{Chen_Wang_2011}

\bibitem[{Chintzoglou et~al.(2019)Chintzoglou, Zhang, Cheung, and
  Kazachenko}]{Chintzoglou_2019}
Chintzoglou, G., Zhang, J., Cheung, M. C.~M., and Kazachenko, M. (2019).
\newblock {The Origin of Major Solar Activity: Collisional Shearing between
  Nonconjugated Polarities of Multiple Bipoles Emerging within Active Regions}.
\newblock \emph{ApJ} 871, 67.
\newblock \doi{10.3847/1538-4357/aaef30}
\bibAnnoteFile{Chintzoglou_2019}

\bibitem[{Falconer et~al.(2002)Falconer, Moore, and Gary}]{Falconer_2002}
Falconer, D.~A., Moore, R.~L., and Gary, G.~A. (2002).
\newblock {Correlation of the Coronal Mass Ejection Productivity of Solar
  Active Regions with Measures of Their Global Nonpotentiality from Vector
  Magnetograms: Baseline Results}.
\newblock \emph{ApJ} 569, 1016--1025.
\newblock \doi{10.1086/339161}
\bibAnnoteFile{Falconer_2002}

\bibitem[{Falconer et~al.(2006)Falconer, Moore, and Gary}]{Falconer_etal_2006}
Falconer, D.~A., Moore, R.~L., and Gary, G.~A. (2006).
\newblock {Magnetic Causes of Solar Coronal Mass Ejections: Dominance of the
  Free Magnetic Energy over the Magnetic Twist Alone}.
\newblock \emph{ApJ} 644, 1258--1272.
\newblock \doi{10.1086/503699}
\bibAnnoteFile{Falconer_etal_2006}

\bibitem[{Falconer et~al.(2008)Falconer, Moore, and Gary}]{Falconer_2008}
Falconer, D.~A., Moore, R.~L., and Gary, G.~A. (2008).
\newblock {Magnetogram Measures of Total Nonpotentiality for Prediction of
  Solar Coronal Mass Ejections from Active Regions of Any Degree of Magnetic
  Complexity}.
\newblock \emph{ApJ} 689, 1433--1442.
\newblock \doi{10.1086/591045}
\bibAnnoteFile{Falconer_2008}

\bibitem[{Georgoulis and Rust(2007)}]{Georgoulis_Rust_2007}
Georgoulis, M.~K. and Rust, D.~M. (2007).
\newblock {Quantitative Forecasting of Major Solar Flares}.
\newblock \emph{ApJ} 661, L109--L112.
\newblock \doi{10.1086/518718}
\bibAnnoteFile{Georgoulis_Rust_2007}

\bibitem[{Green et~al.(2011)Green, Kliem, and Wallace}]{Green_2011}
Green, L.~M., Kliem, B., and Wallace, A.~J. (2011).
\newblock {Photospheric flux cancellation and associated flux rope formation
  and eruption}.
\newblock \emph{A{\&}A} 526, A2.
\newblock \doi{10.1051/0004-6361/201015146}
\bibAnnoteFile{Green_2011}

\bibitem[{Kliem and T{\"{o}}r{\"{o}}k(2006)}]{Kliem_2006}
Kliem, B. and T{\"{o}}r{\"{o}}k, T. (2006).
\newblock {Torus Instability}.
\newblock \emph{PhRvL} 96, 255002.
\newblock \doi{10.1103/PhysRevLett.96.255002}
\bibAnnoteFile{Kliem_2006}

\bibitem[{K{\"{u}}nzel(1960)}]{Kunzel_1960}
K{\"{u}}nzel, H. (1960).
\newblock {Die Flare-H{\"{a}}ufigkeit in Fleckengruppen unterschiedlicher
  Klasse und magnetischer Struktur}.
\newblock \emph{Astronomische Nachrichten} 285, 271
\bibAnnoteFile{Kunzel_1960}

\bibitem[{Leka and Barnes(2003)}]{Leka_Barnes_2003a}
Leka, K.~D. and Barnes, G. (2003).
\newblock {Photospheric Magnetic Field Properties of Flaring versus
  Flare‐quiet Active Regions. I. Data, General Approach, and Sample Results}.
\newblock \emph{ApJ} 595, 1277--1295.
\newblock \doi{10.1086/377511}
\bibAnnoteFile{Leka_Barnes_2003a}

\bibitem[{Leka and Barnes(2007)}]{Leka2007}
Leka, K.~D. and Barnes, G. (2007).
\newblock {Photospheric Magnetic Field Properties of Flaring versus
  Flare‐quiet Active Regions. IV. A Statistically Significant Sample}.
\newblock \emph{ApJ} 656, 1173--1186.
\newblock \doi{10.1086/510282}
\bibAnnoteFile{Leka2007}

\bibitem[{Lemen et~al.(2012)Lemen, Title, Akin, Boerner, Chou, Drake
  et~al.}]{Lemen2012}
Lemen, J.~R., Title, A.~M., Akin, D.~J., Boerner, P.~F., Chou, C., Drake,
  J.~F., et~al. (2012).
\newblock {The Atmospheric Imaging Assembly (AIA) on the Solar Dynamics
  Observatory (SDO)}.
\newblock \emph{SoPh} 275, 17--40.
\newblock \doi{10.1007/s11207-011-9776-8}
\bibAnnoteFile{Lemen2012}

\bibitem[{Li and Zhang(2009)}]{Li_2009}
Li, L. and Zhang, J. (2009).
\newblock {STATISTICS OF FLARES SWEEPING ACROSS SUNSPOTS}.
\newblock \emph{ApJ} 706, 17--21.
\newblock \doi{10.1088/0004-637X/706/1/L17}
\bibAnnoteFile{Li_2009}

\bibitem[{Liu et~al.(2015)Liu, Wang, Shen, Liu, Pan, and Wang}]{Liu_2015}
Liu, J., Wang, Y., Shen, C., Liu, K., Pan, Z., and Wang, S. (2015).
\newblock {A SOLAR CORONAL JET EVENT TRIGGERS A CORONAL MASS EJECTION}.
\newblock \emph{ApJ} 813, 115.
\newblock \doi{10.1088/0004-637X/813/2/115}
\bibAnnoteFile{Liu_2015}

\bibitem[{Liu et~al.(2018)Liu, Cheng, Wang, Zhou, Guo, and Cui}]{Liu_2018d}
Liu, L., Cheng, X., Wang, Y., Zhou, Z., Guo, Y., and Cui, J. (2018).
\newblock {Rapid Buildup of a Magnetic Flux Rope during a Confined X2.2 Class
  Flare in NOAA AR 12673}.
\newblock \emph{ApJ} 867, L5.
\newblock \doi{10.3847/2041-8213/aae826}
\bibAnnoteFile{Liu_2018d}

\bibitem[{Liu et~al.(2016)Liu, Wang, Wang, Shen, Ye, Liu et~al.}]{Lliu_2016}
Liu, L., Wang, Y., Wang, J., Shen, C., Ye, P., Liu, R., et~al. (2016).
\newblock {WHY IS A FLARE-RICH ACTIVE REGION CME-POOR?}
\newblock \emph{ApJ} 826, 119.
\newblock \doi{10.3847/0004-637X/826/2/119}
\bibAnnoteFile{Lliu_2016}

\bibitem[{Liu et~al.(2021)Liu, Wang, Zhou, and Cui}]{Liu_2021}
Liu, L., Wang, Y., Zhou, Z., and Cui, J. (2021).
\newblock {The Source Locations of Major Flares and CMEs in Emerging Active
  Regions}.
\newblock \emph{ApJ} 909, 142.
\newblock \doi{10.3847/1538-4357/abde37}
\bibAnnoteFile{Liu_2021}

\bibitem[{Moore et~al.(2001)Moore, Sterling, Hudson, and Lemen}]{Moore2001a}
Moore, R.~L., Sterling, A.~C., Hudson, H.~S., and Lemen, J.~R. (2001).
\newblock {Onset of the Magnetic Explosion in Solar Flames and Coronal Mass
  Ejections}.
\newblock \emph{ApJ} 552, 833--848.
\newblock \doi{10.1086/320559}
\bibAnnoteFile{Moore2001a}

\bibitem[{Pesnell et~al.(2012)Pesnell, Thompson, and Chamberlin}]{Pesnell_2012}
Pesnell, W.~D., Thompson, B.~J., and Chamberlin, P.~C. (2012).
\newblock {The Solar Dynamics Observatory (SDO)}.
\newblock \emph{SoPh} 275, 3--15.
\newblock \doi{10.1007/s11207-011-9841-3}
\bibAnnoteFile{Pesnell_2012}

\bibitem[{Scherrer et~al.(2012)Scherrer, Schou, Bush, Kosovichev, Bogart,
  Hoeksema et~al.}]{Scherrer_2012}
Scherrer, P.~H., Schou, J., Bush, R.~I., Kosovichev, A.~G., Bogart, R.~S.,
  Hoeksema, J.~T., et~al. (2012).
\newblock {The Helioseismic and Magnetic Imager (HMI) Investigation for the
  Solar Dynamics Observatory (SDO)}.
\newblock \emph{SoPh} 275, 207--227.
\newblock \doi{10.1007/s11207-011-9834-2}
\bibAnnoteFile{Scherrer_2012}

\bibitem[{Schmieder et~al.(2014)Schmieder, Archontis, and
  Pariat}]{Schmieder_2014}
Schmieder, B., Archontis, V., and Pariat, E. (2014).
\newblock {Magnetic Flux Emergence Along the Solar Cycle}.
\newblock \emph{SSRv} 186, 227--250.
\newblock \doi{10.1007/s11214-014-0088-9}
\bibAnnoteFile{Schmieder_2014}

\bibitem[{Schrijver(2007)}]{schrijver2007characteristic}
Schrijver, C.~J. (2007).
\newblock {A characteristic magnetic field pattern associated with all major
  solar flares and its use in flare forecasting}.
\newblock \emph{ApJL} 655, L117
\bibAnnoteFile{schrijver2007characteristic}

\bibitem[{Schrijver(2009)}]{Schrijver_2009}
Schrijver, C.~J. (2009).
\newblock {Driving major solar flares and eruptions: A review}.
\newblock \emph{AdSpR} 43, 739--755.
\newblock \doi{10.1016/j.asr.2008.11.004}
\bibAnnoteFile{Schrijver_2009}

\bibitem[{Strous and Zwaan(1999)}]{Strous_1999}
Strous, L.~H. and Zwaan, C. (1999).
\newblock {Phenomena in an Emerging Active Region. II. Properties of the
  Dynamic Small‐Scale Structure}.
\newblock \emph{ApJ} 527, 435--444.
\newblock \doi{10.1086/308071}
\bibAnnoteFile{Strous_1999}

\bibitem[{Sun et~al.(2015)Sun, Bobra, Hoeksema, Liu, Li, Shen
  et~al.}]{Sunxd_2015}
Sun, X., Bobra, M.~G., Hoeksema, J.~T., Liu, Y., Li, Y., Shen, C., et~al.
  (2015).
\newblock {WHY IS THE GREAT SOLAR ACTIVE REGION 12192 FLARE-RICH BUT CME-POOR?}
\newblock \emph{ApJ} 804, L28.
\newblock \doi{10.1088/2041-8205/804/2/L28}
\bibAnnoteFile{Sunxd_2015}

\bibitem[{Toriumi et~al.(2016)Toriumi, Schrijver, Harra, Hudson, and
  Nagashima}]{Toriumi_2016}
Toriumi, S., Schrijver, C.~J., Harra, L.~K., Hudson, H., and Nagashima, K.
  (2016).
\newblock {MAGNETIC PROPERTIES OF SOLAR ACTIVE REGIONS THAT GOVERN LARGE SOLAR
  FLARES AND ERUPTIONS}.
\newblock \emph{ApJ} 834, 56.
\newblock \doi{10.3847/1538-4357/834/1/56}
\bibAnnoteFile{Toriumi_2016}

\bibitem[{T{\"{o}}r{\"{o}}k et~al.(2004)T{\"{o}}r{\"{o}}k, Kliem, and
  Titov}]{Torok_2004}
T{\"{o}}r{\"{o}}k, T., Kliem, B., and Titov, V.~S. (2004).
\newblock {Ideal kink instability of a magnetic loop equilibrium}.
\newblock \emph{A{\&}A} 413, L27--L30.
\newblock \doi{10.1051/0004-6361:20031691}
\bibAnnoteFile{Torok_2004}

\bibitem[{van Ballegooijen and Martens(1989)}]{VanBallegooijen_1989}
van Ballegooijen, A.~A. and Martens, P. C.~H. (1989).
\newblock {Formation and eruption of solar prominences}.
\newblock \emph{ApJ} 343, 971.
\newblock \doi{10.1086/167766}
\bibAnnoteFile{VanBallegooijen_1989}

\bibitem[{van Driel-Gesztelyi and Green(2015)}]{Van_2015}
van Driel-Gesztelyi, L. and Green, L.~M. (2015).
\newblock {Evolution of Active Regions}.
\newblock \emph{LRSP} 12, 1.
\newblock \doi{10.1007/lrsp-2015-1}
\bibAnnoteFile{Van_2015}

\bibitem[{Wang et~al.(2011)Wang, Chen, Gui, Shen, Ye, and Wang}]{Wang_2011c}
Wang, Y., Chen, C., Gui, B., Shen, C., Ye, P., and Wang, S. (2011).
\newblock {Statistical study of coronal mass ejection source locations:
  Understanding CMEs viewed in coronagraphs}.
\newblock \emph{JGRA} 116, A04104.
\newblock \doi{10.1029/2010JA016101}
\bibAnnoteFile{Wang_2011c}

\bibitem[{Welsch et~al.(2004)Welsch, Fisher, Abbett, and Regnier}]{Welsch_2004}
Welsch, B.~T., Fisher, G.~H., Abbett, W.~P., and Regnier, S. (2004).
\newblock {ILCT: Recovering Photospheric Velocities from Magnetograms by
  Combining the Induction Equation with Local Correlation Tracking}.
\newblock \emph{ApJ} 610, 1148--1156.
\newblock \doi{10.1086/421767/FULLTEXT/}
\bibAnnoteFile{Welsch_2004}

\bibitem[{Yan and Qu(2007)}]{Yan_2007}
Yan, X.~L. and Qu, Z.~Q. (2007).
\newblock {Rapid rotation of a sunspot associated with flares}.
\newblock \emph{A{\&}A} 468, 1083--1088.
\newblock \doi{10.1051/0004-6361:20077064}
\bibAnnoteFile{Yan_2007}

\bibitem[{Yan et~al.(2008)Yan, Qu, and Kong}]{Yan_2008}
Yan, X.~L., Qu, Z.~Q., and Kong, D.~F. (2008).
\newblock {Relationship between rotating sunspots and flare productivity}.
\newblock \emph{MNRAS} 391, 1887--1892.
\newblock \doi{10.1111/j.1365-2966.2008.14002.x}
\bibAnnoteFile{Yan_2008}

\bibitem[{Yan et~al.(2009)Yan, Qu, Xu, Xue, and Kong}]{Yan_2009}
Yan, X.-L., Qu, Z.-Q., Xu, C.-L., Xue, Z.-K., and Kong, D.-F. (2009).
\newblock {The causality between the rapid rotation of a sunspot and an X3.4
  flare}.
\newblock \emph{Research in Astronomy and Astrophysics} 9, 596--602.
\newblock \doi{10.1088/1674-4527/9/5/010}
\bibAnnoteFile{Yan_2009}

\bibitem[{Yan et~al.(2015)Yan, Xue, Pan, Wang, Xiang, Kong et~al.}]{Yanxl_2015}
Yan, X.~L., Xue, Z.~K., Pan, G.~M., Wang, J.~C., Xiang, Y.~Y., Kong, D.~F.,
  et~al. (2015).
\newblock {THE FORMATION AND MAGNETIC STRUCTURES OF ACTIVE-REGION FILAMENTS
  OBSERVED BY NVST, SDO , AND HINODE}.
\newblock \emph{ApJS} 219, 17.
\newblock \doi{10.1088/0067-0049/219/2/17}
\bibAnnoteFile{Yanxl_2015}

\bibitem[{Zhang et~al.(2007)Zhang, Li, and Song}]{Zhang_2007}
Zhang, J., Li, L., and Song, Q. (2007).
\newblock {INTERACTION BETWEEN A FAST ROTATING SUNSPOT AND EPHEMERAL REGIONS AS
  THE ORIGIN OF THE MAJOR SOLAR EVENT ON 2006 DECEMBER 13}.
\newblock \emph{ApJ} 662, 35--38
\bibAnnoteFile{Zhang_2007}

\bibitem[{Zirin and Margaret(1987)}]{Zirin_1987}
Zirin, H. and Margaret, A.~L. (1987).
\newblock {DELTA SPOTS AND GREAT FLARES}.
\newblock \emph{SoPh} 113, 267--283
\bibAnnoteFile{Zirin_1987}

\end{thebibliography}


\section*{Figures}


\end{document}